\documentclass[aps,prb,onecolumn,superscriptaddress]{revtex4}
\usepackage{graphics}
\usepackage{longtable}
\usepackage[epsfig]{graphicx}
\usepackage{epsfig}
\usepackage{amsmath}
\usepackage{float}
\begin{document}
	
	\title{Near-perfect spin filtering and negative differential resistance in an Fe(II)S complex}

	\author{Sherif Abdulkader Tawfik}
	\email{sherif.abbas@uts.edu.au}
	\affiliation{School of Physics, The University of Sydney, New South Wales, 2006, Australia}

	\author{Leigh Weston}
	\affiliation{Materials Department, University of California, Santa Barbara, CA, USA}

	\author{X. Y. Cui}
	\affiliation{Australian Institute for Nanoscale Science and Technology, and School of Aerospace, Mechanical and Mechatronic Engineering, The University of Sydney, New South Wales, 2006, Australia.}

	\author{Simon Ringer}
	\affiliation{Australian Institute for Nanoscale Science and Technology, and School of Aerospace, Mechanical and Mechatronic Engineering, The University of Sydney, New South Wales, 2006, Australia.}

	\author{Catherine Stampfl}
	\affiliation{School of Physics, The University of Sydney, New South Wales, 2006, Australia}
	
\begin{abstract}
Density functional theory and nonequilibrium Green's function calculations have been used to explore spin-resolved transport through the high-spin state of an iron(II)sulfur single molecular magnet. Our results show that this molecule exhibits near-perfect spin filtering, where the spin-filtering efficiency is above 99\%, as well as significant negative differential resistance centered at a low bias voltage. The rise in the spin-up conductivity up to the bias voltage of 0.4 V is dominated by a conductive lowest unoccupied molecular orbital, and this is accompanied by a slight increase in the magnetic moment of the Fe atom. The subsequent drop in the spin-up conductivity is because the conductive channel moves to the highest occupied molecular orbital which has a lower conductance contribution. This is accompanied by a drop in the magnetic moment of the Fe atom. These two exceptional properties, and the fact that the onset of negative differential resistance occurs at low bias voltage, suggests the potential of the molecule in nanoelectronic and nanospintronic applications.
\end{abstract}
\maketitle

The generation of highly spin-polarised currents (that is, spin filtering) remains one of the most important targets in the field of spintronics. Since the first experimental demonstration of spin filtering discovered by M\"{u}ller \textit{et al.} in 1972,\cite{[1]} control of the electron spin in electrical transport in solid-state spintronics has led to important applications in data storage and magnetic sensing.\cite{[2]} In the emerging field of molecular spintronics,\cite{[3], [4]} spin-coherence may be preserved over time and distance much longer than in conventional metals or semiconductors. There are in principle two approaches for developing molecule-based spin-filtering devices. The first option is to place a diamagnetic molecular bridge between two ferromagnetic electrodes.\cite{[5]} The second option is to couple a paramagnetic molecule, also known as a single molecular magnet (SMM), to two diamagnetic electrodes.\cite{[6], [7]} The latter approach does not require magnetic electrodes and can be turned on and off by adjusting an electrostatic gate voltage. Quantum transport through a SMM that connects source and drain electrodes depends strongly on the alignment of molecular energy levels with respect to the chemical potentials at both electrodes. In this regard, the intrinsic molecular orbitals of SMMs, are the key ingredient in determining the spin-resolved transport properties.\cite{[7],[8]}

Among the SMMs investigated in the literature, some Fe-containing SMMs are known to preserve their intrinsic magnetic properties after adsorption on an Au\cite{sf18} or Ru surface.\cite{sf18-1} In particular, Fe(II)$-$S complexes are a common component of biological systems and have critical roles in a broad range of biological processes.\cite{1} These clusters are well known for electron-transfer and redox reactions \cite{[10]}. Very recently, a rod-shaped Fe(II)$-$S complex of proteins has been proposed to serve as a biocompass, which can use the Earth's weak magnetic field to navigate or orient for many organisms \cite{[11]}.

\begin{figure}
\includegraphics[width=90mm]{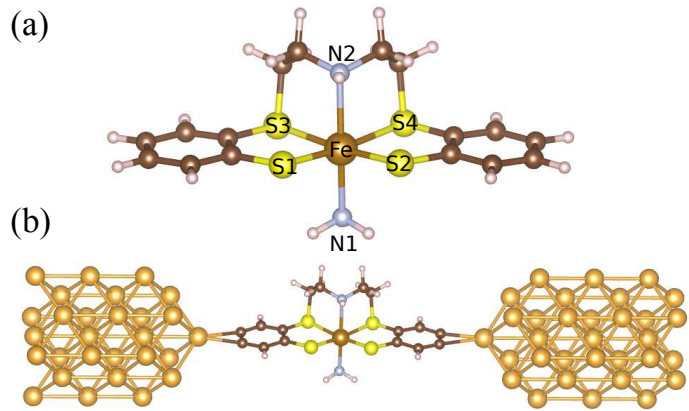}
  \caption{(a) The atomic geometry of the Fe(II)S molecule (in the HS state). (b) The Fe(II)S complex without the four terminal H atoms and between two Au electrodes. The golden sphere represents Fe, brown spheres for C, white spheres for H, yellow spheres for S and cyan spheres for Au. The Fe$-$S1/2 bond length is 2.33 {\AA}; Fe$-$S3 is 2.59 {\AA}; Fe$-$S4 is 2.58 {\AA}; Fe$-$N1 is 2.24 {\AA}; Fe$-$N2 is 2.29 {\AA}.}
  \label{fig:fig1}
\end{figure}

Fe(II) complexes have been shown to exhibit dramatic conductance difference between the high-spin (HS) and low-spin (LS) states at low voltages,\cite{13,14} and recent studies \cite{sf15,sf19} showed that an Fe complexes, in addition to their high spin-filtering performance, also exhibits negative differential resistance (NDR). Such behavior was attributed to the strong coupling between the electrode and the Fe $3d$ orbitals through organic ligands.\cite{sf19} For the purpose of this paper, we study the [Fe(N$_{\rm H}$)S$_4$]NH$_3$ molecule (cf. Fig. \ref{fig:fig1}(a)) (henceforth denoted as Fe(II)S for short) whose ground state is a HS state as demonstrated experimentally\cite{Sellmann} and theoretically.\cite{4}

NDR is characterized by a decreasing current with increasing bias in some specific bias range. NDR has been reported in a number of nanostructures,\cite{15,16,17} and allows for fast switching in certain electronic devices such as fast switches, oscillators, and other high-frequency electronic devices.\cite{ndrjap} The combined NDR-spin-filtering effects have been reported in numerous studies for various systems\cite{sf15,18} but to date has not been reported for Fe(II) complexes.\cite{13}

In this paper, we report the possibility of near-perfect spin-filtering and a significant NDR effect in the Fe(II)S molecule when connected to two Au electrodes (cf. Fig. \ref{fig:fig1}(b)). Spin-polarization of the current is assessed by the spin-filtering efficiency (SFE), defined as $\left|I_{\alpha}-I_{\beta}\right|/\left|I_{\alpha}+I_{\beta}\right|$, where $I_{\alpha}$ and $I_{\beta}$ refer to the current in the spin-up and spin-down channels, respectively. We report a SFE above 99.9\% in the bias region 0.0 V to 0.3 V.

The atomic geometry of the Fe(II)S molecule is shown in Fig. \ref{fig:fig1}(a). The Fe atom is bonded to four S atoms, and to two N atoms (N-H bonds form a tent-like structure above Fe, and NH$_{3}$ below). In the transport simulation, the molecule is connected to the Au electrodes via benzene groups. Note that we remove the two terminal H atoms in the benzene groups on each side in order to enhance the bonding between electrodes and the $\pi$ orbitals of the molecule, thus enhancing the conductance across the electrode-ligand junction. Direct metal-benzyne (C$_{6}$H$_{4}$) adsorption (without the usual thiol group anchoring) was experimentally found to result from the dissociation of benzene.\cite{C6H4} This finding supports the model in which the Fe(II)S complex with 2H dissociated from the two side benzene groups is used as a molecular junction. The gold electrodes extend semi-infinitely in both directions along the $z$ axis, and we use a vacuum separation of 20 {\AA} between repeated images along the transverse $x$ and $y$ axes in order to minimize any possible interactions between the complex and its mirror images. We utilize a Au(100)-based nanowire electrode structure as shown in Fig. \ref{fig:fig1}(b). The electrodes are created from the bulk gold structure, where the number of atoms in each layer (apart from the two layers of the tip) along the electrode ($z$) axis is 5, 4, 5, 4, as has been previously used for an Al electrode,\cite{Jeremy} an Ag electrode\cite{Ag} and an Au electrode.\cite{jcptawfik} The left electrode has 28 Au atoms, and the right electrode has 32 Au atoms. The repeating unit cell along the $z$ direction in the electrode has 18 atoms (so, there are 10 Au atoms included in the scattering region from the right electrode and 14 Au atoms from the left electrode). 

The Fe(II)S molecule was firstly relaxed in vacuum using density functional theory (DFT) using the generalized gradient approximation (GGA) of Perdew, Burke and Ernzerhof (PBE)\cite{PBE} and the Heyd-Scuseria-Ernzerhof (HSE) hybrid functional.\cite{HSE} The valence electrons are separated from the core by use of projector-augmented wave pseudopotentials (PAW) \cite{BlochlPRB1994} as implemented in the VASP package \cite{KressePRB1996}. For the present calculations C $2s^22p^2$, N $2s^22p^3$, O $2s^22p^4$, S $3s^23p^4$ and Fe $4s^23d^6$ electrons are treated as valence. The energy cut-off for the plane wave basis set is 500 eV. The 43 atom molecule is simulated by placing it inside of a box with dimensions 25 \AA$\times$17 \AA$\times$17 \AA. During the structural energy minimization the internal coordinates are allowed to relax until all of the forces are less than 0.01 eV/\AA. The equilibrium Au$-$C$_{6}$H$_{4}$ separation distance is 2.23 \AA. We have performed the current-voltage characteristics for Au$-$C$_{6}$H$_{4}$ separations of 2.2 \AA, 2.5 \AA and 2.75 \AA, and the results are quantitatively close. We chose the electrode-complex distance to be 2.5 {\AA}, which is within the range of bond distances reported in Ref. \citenum{acsnano}.

The transport calculations are performed using the nonequilibrium Green's function (NEGF) method as implemented in the TRANSIESTA code.\cite{TRANSIESTA} Double-$\zeta$ is used for Fe, S, N, C and H while single-$\zeta$ is used for Au. When a molecule is placed between electrodes (Fig. \ref{fig:fig1}(b)), the current $I_{\sigma}$ through the scattering region, for the spin channel $\sigma$, as a function of the bias voltage $V_{\textrm{\tiny{bias}}}$ across the device can be estimated using the Landauer-Buttiker formula,\cite{Landauer} 

\begin{equation}
I_{\sigma}(V_{\textrm{\tiny{bias}}})= \frac{2e}{h}\int^{-\infty}_{\infty}T_{\sigma}(E,V_{\textrm{\tiny{bias}}})\times \left[f_{L}(E-\mu_{L})-f_{R}(E-\mu_{R})\right]dE, 
	\label{LB}
\end{equation}

\noindent where $L$ and $R$ denote left and right electrodes, respectively, $T_{\sigma}(E,V_{\textrm{\tiny{bias}}})$ is the transmission function for spin channel $\sigma$, which is a function of the energy ($E$) and $V_{\textrm{\tiny{bias}}}$, the voltage applied across the electrodes. $f_{L/R}$ is the Fermi-Dirac distribution function and $\mu_{L/R}$ is the electrochemical potential. $T_{\sigma}(E,V_{\textrm{\tiny{bias}}})$ is the trace of the square of the transmission amplitude $\textbf{t}_{\sigma}$, and it takes the form,

\begin{equation}
	T_{\sigma}(E,V_{\textrm{\tiny{bias}}})= \textrm{Tr}\left[\textbf{t}_{\sigma}^{\dagger}\textbf{t}_{\sigma}\right]=\textrm{Tr}\left[\Gamma_{R\sigma}G_{\sigma}\Gamma_{L\sigma}G_{\sigma}^{\dagger}\right],
	\label{eq1}
\end{equation}
 
\noindent where $\Gamma_{R/L\sigma}$ is the imaginary part of the self-energy, and $G_{\sigma}$ is the Green's function of the scattering region.

\begin{figure}
	\includegraphics[width=90mm]{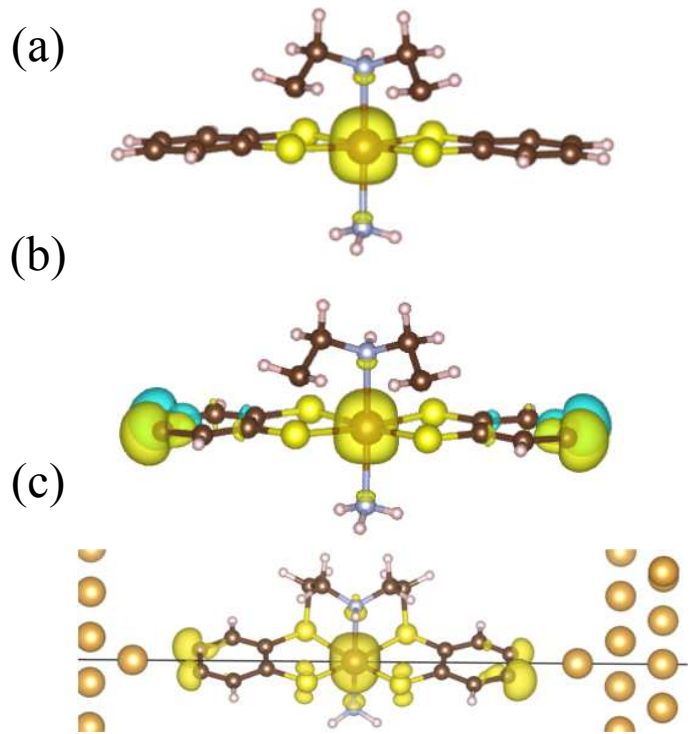}
	\caption{The spin density of the (a) Fe(II)S molecule, (b) the Fe(II)S molecule after removing the four terminal H atoms, and (c) the gold-Fe(II)S-gold system. The golden spheres represents the Fe atom, brown spheres for C, white spheres for H, yellow spheres for S and cyan spheres for Au.}
	\label{fig:SpinDensity}
\end{figure}

Experimentally, the HS state is the ground state of the Fe(II)S molecule.\cite{4,Sellmann} However, performing a DFT calculation using the GGA will result in an energy difference between the HS and LS state of $E_{HS-LS}=E_{HS}-E_{LS}=+0.79$ eV using VASP, that is, it predicts the LS state is energetically more favorable than the HS state. We note that using SIESTA, with the GGA, the atomic structure is consistent with VASP, and the LS state is also predicted to be favorable over the HS state. The experimental behavior is reproduced when we use the HSE hybrid functional as implemented in VASP. In particular, our calculations show that $E_{HS-LS}$ depends linearly on the proportion of exact exchange (EXX) included,\cite{Leigh} where a value EXX=0.0 gives $E_{HS-LS}=+0.79$ eV, EXX=0.125 gives $E_{HS-LS}=+0.13$ eV, EXX=0.15 gives $E_{HS-LS}=-0.016$ eV and EXX=0.2 gives $E_{HS-LS}=-0.26$ eV. The value of EXX=0.15 is consistent with that reported in Ref. \citenum{4}. The total spin moment obtained using EXX=0.15 is 4 $\mu_B$, where there is 3.55 $\mu_B$ on Fe, 0.031/0.052 $\mu_B$ on the S atoms, and 0.22 $\mu_B$ in the interstitial region (see Fig. 2(a) where the spin density is plotted). Interestingly, the total spin moment for the Fe(II)S molecule with 4H atoms removed (2H atoms removed from each end of the molecule, where the calculation was performed self-consistently using the atomic geometry of the full molecule) is also 4 $\mu_B$, with 3.48 $\mu_B$ on Fe, 0.032/0.043 $\mu_B$ on S atoms, and 0.402/-0.386 $\mu_B$ on the unsatuated edge C atoms (with the neighboring C pair being antiferromagnetic) (see Fig. 2(b)). In the LS state (which has zero spin moment), the main difference to the HS state geometry is that the Fe$-$N1 bond length is 2.05 \AA, and the Fe$-$N2 bond length is 2.04 \AA, so they are shorter than the corresponding bond-lengths in the HS state (which are 2.24 \AA and 2.29 \AA, respectively). For our transport calculations, we choose to use the HS atomic structure relaxed using GGA, given that the TranSIESTA implementation does not incorporate hybrid functionals, and that the difference in the atomic structure between the GGA-relaxed and HSE-relaxed geometries is very small. We display the spin-density isoplots in Fig. \ref{fig:SpinDensity} for the gold-Fe(II)S-gold system, calculated using GGA. The corresponding value of the spin-moment is 3.937 $\mu_B$ using GGA, and 4.484 $\mu_B$ using HSE06.


\begin{figure}
	\includegraphics[width=90mm]{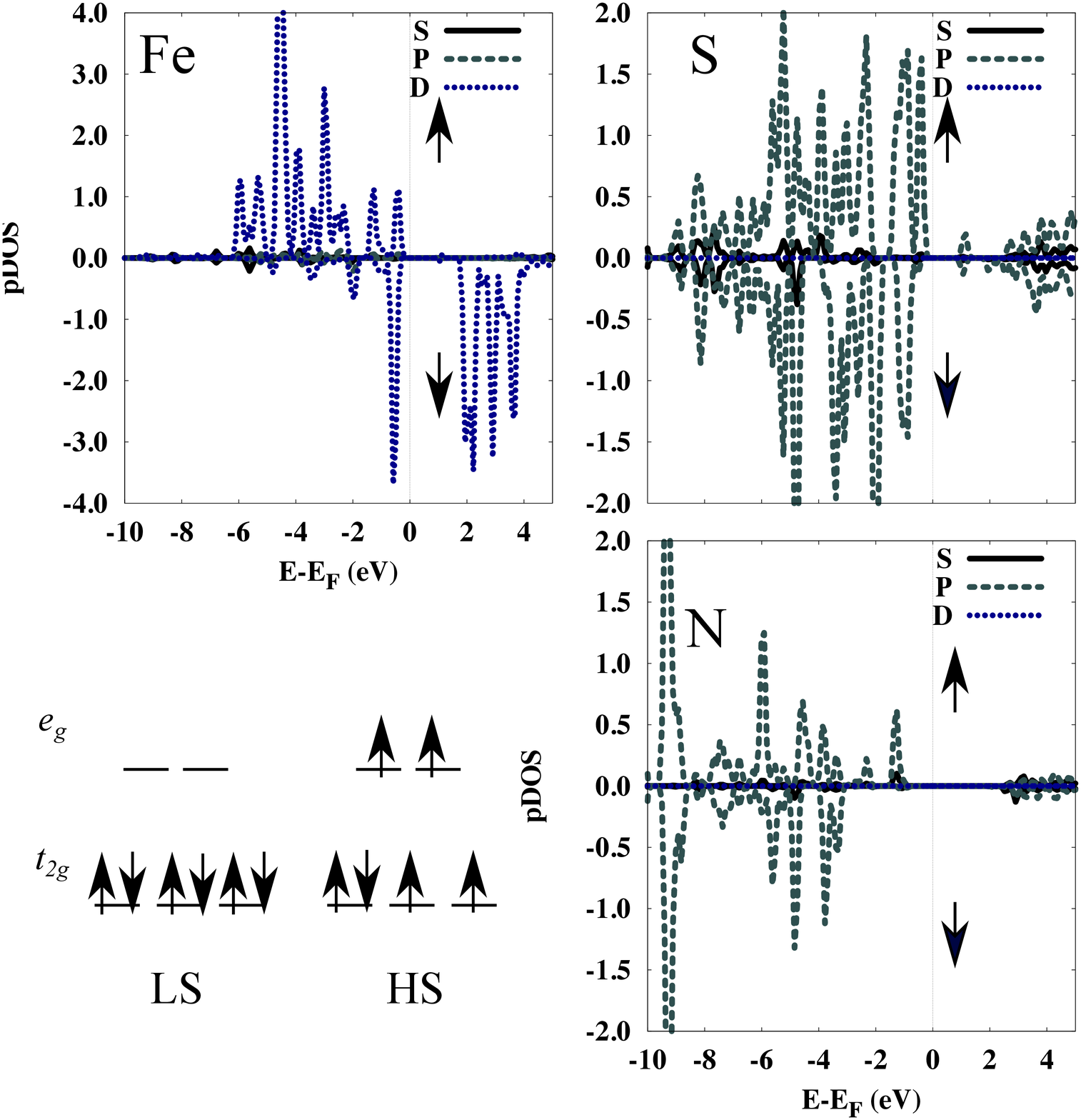}
	\caption{The spin-up $\uparrow$ and spin-dpwn $\downarrow$ pDOS of the Fe(II)S molecule after removing the four terminal H atoms. Lower-left is simple ligand field theory showing the splitting of the Fe $d^6$ state and occupancy.}
	\label{fig:fig_PDOS_Isolated_noH}
\end{figure}

We now consider the electronic structure of the Fe(II)S molecule. We display the pDOS of the isolated Fe(II)S molecule with the four H atoms removed in Fig. \ref{fig:fig_PDOS_Isolated_noH}. These are calculated using VASP (HSE06). In the simple ligand field theory, the Fe(II) atom would have occupancy $d^6$, where all six $d$ electrons may occupy the $t_{2g}$ orbitals or may occupy both $t_{2g}$ and $e_g$ orbitals (illustrated in Fig. \ref{fig:fig_PDOS_Isolated_noH}, lower left panel). This corresponds, respectively, to having a LS ($S=0$) singlet ground state or a HS ($S=2$) ground state. From the pDOS shown in Fig. \ref{fig:fig_PDOS_Isolated_noH}, the HS configuration can be clearly seen, exhibiting largely occupied Fe spin-up states and only one Fe spin-down is state occupied. The S $p$-state has a partially occupied spin-up state at the Fermi level showing hybridization with the Fe $d$-state. Removing the H atoms causes the partially occupied peak in the $3d$ orbital of Fe to become fully occupied and shift down in energy.

\begin{figure}
	\includegraphics[width=90mm]{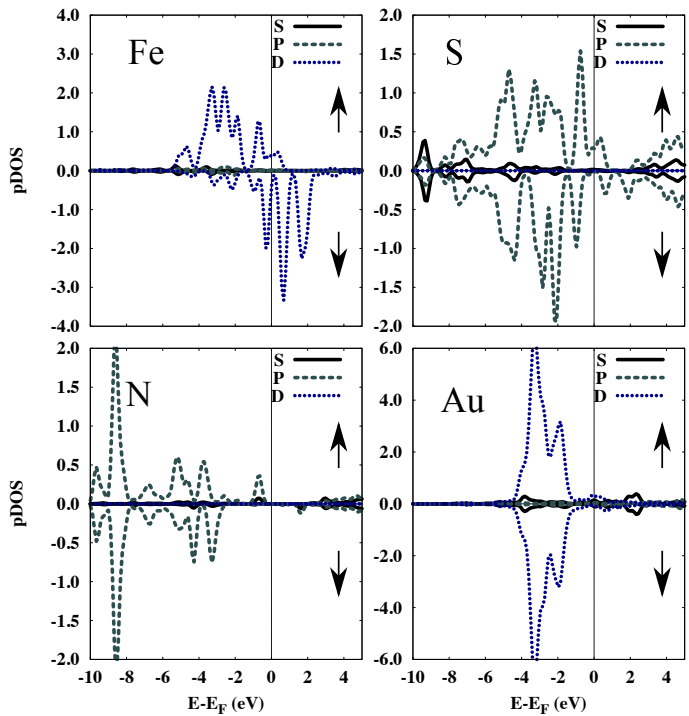}
	\caption{The spin-up $\uparrow$ and spin-dpwn $\downarrow$ pDOS of the gold-Fe(II)S-gold system.}
	\label{fig:fig_PDOS_TotalSystem}
\end{figure}

The partial density of states (pDOS) of the gold-Fe(II)S-gold is displayed in Fig. \ref{fig:fig_PDOS_TotalSystem}, calculated using VASP (GGA). The behavior of the pDOS is consistent with the spin filtering behavior (discussed below): there is considerable pDOS at the Fermi level in both the spin-up states of Fe $3d$ and S $3p$, as well as a smaller combination of Au $5d$ states, while there is no such pDOS in the spin-down state at the Fermi level. This analysis shows the important role of the electrode in the spin-filtering behavior of Fe(II)S. The N atoms does not contribute to the spin-filtering behavior, given that its $2p$ orbital is fully occupied, as in the free molecule (cf. Fig. \ref{fig:fig_PDOS_Isolated_noH}).

\begin{figure}
\includegraphics[width=90mm]{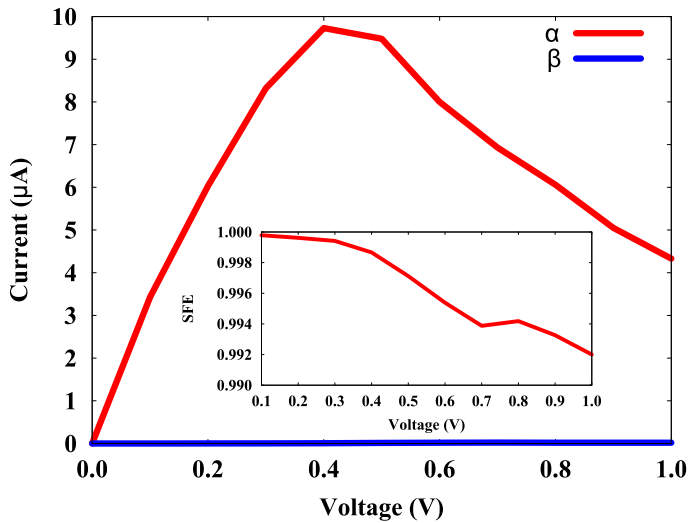}
  \caption{The current-voltage characteristics (IVC) of the gold-Fe(II)S-gold system, where the red line denotes the $I-V$ of the spin-up channel, and the blue line denotes the IVC of the spin-down channel. The inset shows the calculated spin-filtering efficiency (SFE).}
  \label{fig:fig3}
\end{figure}

The calculated spin-resolved current-voltage characteristics (IVC) of the complex is shown in Fig. \ref{fig:fig3}. The IVC shows that the current flow is dominated almost entirely by the spin-up channel, whereas the spin-down current remains close to 0 for voltage ranging from 0.0 to 1.0 V. The inset shows that the SFE is above 99.9\% for the bias voltage range 0.0 V to 0.3 V, and 99.2\% for the whole voltage range considered. Moreover, the system experiences a large NDR effect starting at 0.4 V. We considered slight variations in the Fe(II)S$-$electrode distance and found that they do not impact the maximum SFE. That is, we have calculated the IVC for Fe(II)S$-$electrode distances of 2.2 {\AA}, and 2.75 {\AA}, and the resulting maximum SFE was still 99.9\% while the magnitude of the current, as well as the NDR position, was not significantly affected.

\begin{figure}
\includegraphics[width=90mm]{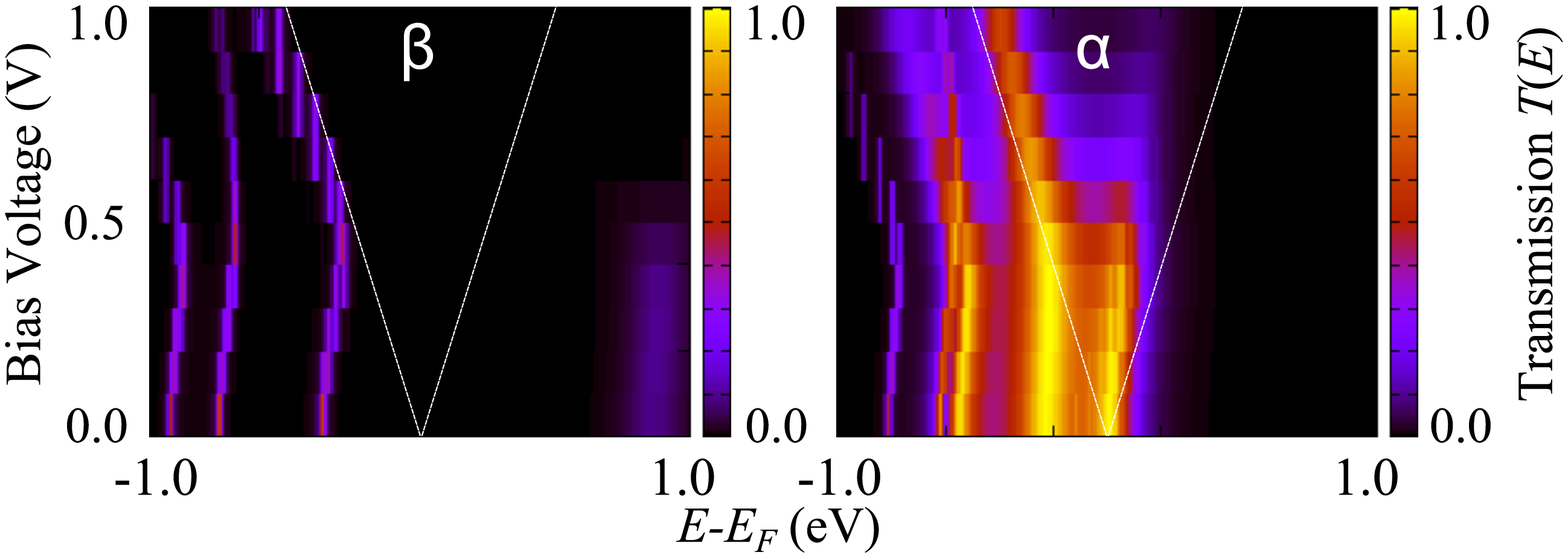}
  \caption{The transmission function, $T(E,V_{\textrm{\tiny{\rm bias}}})$ as a function of energy $E$ and bias voltage $V_{\textrm{\tiny{\rm bias}}}$ for the spin-down (left) and spin-up (right) channels.}
  \label{fig:fig4}
\end{figure}

To rationalize the strong spin-filtering and NDR behaviours, we study the transmission function of both spin channels as a function of energy and bias voltage, as shown in Fig.~\ref{fig:fig4}. The bias window is depicted as two white lines intersecting at the origin $E-E_{F}=0$. The figure shows that there is a large difference in the transmission function and hence the conductance between the two spin channels. While the transmission function of the spin-up channel is characterized by large transmission within the bias window for the whole bias voltage range, the transmission function in the spin-down channel does not show any significant transmission within the bias window. For the spin-up channel, the region of large transmission expands as $V_{\textrm{\tiny{\rm bias}}}$ approaches 0.4 V, and starts decaying as $V_{\textrm{\tiny{\rm bias}}}$ exceeds 0.4 V. Thus, the current will decrease as the bias voltage exceeds 0.4 V.

\begin{figure}
\includegraphics[width=90mm]{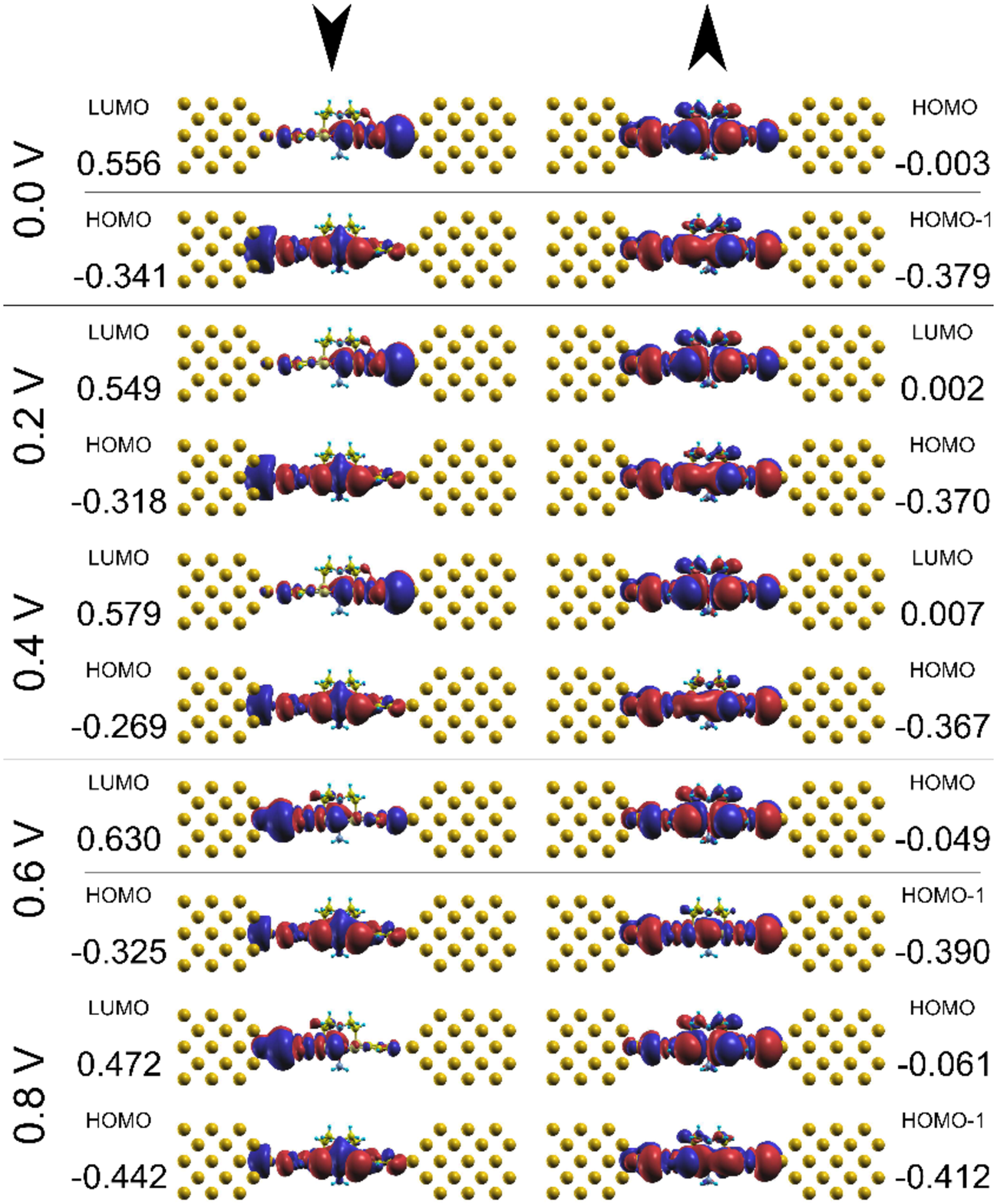}
  \caption{The MPSH orbital isographs for isolevel 0.01 for the spin-up and spin-down channels at voltage bias 0.0 V, 0.2 V, 0.4 V, 0.6 V and 0.8 V. Current flows from left to right. For each isograph we provide the eigenchannel energy (in units of eV) as well as the orbital label (HOMO, LUMO or HOMO-1). Note that we have selected orbitals that lie within, as well as on the boundary of, the bias window.}
  \label{fig:fig5}
\end{figure}

In order to understand the molecular orbitals responsible for the transmission discussed above, we turn to the analysis of the Molecular Projected Self-Consistent Hamiltonian (MPSH). The MPSH orbitals are the eigenvectors of the transmission matrix, projected on the system's molecular orbitals. They provide a simplified picture of electron transport, and give a direct spatially resolved picture of the orbitals involved in the transport.\cite{inelastica,st2,st3} MPSH orbitals also enable us to analyze transmission properties by studying two features: the degree of localization or delocalization of the orbital (by visual inspection) and whether the eigenenergy of the MPSH orbital is inside or outside of the bias window (defined as $[-V_{\textrm{\tiny{\rm bias}}}/2,+V_{\textrm{\tiny{\rm bias}}}/2]$). Figure~\ref{fig:fig5} shows the MPSH orbitals for bias voltages $V_{\textrm{\tiny{bias}}}=$ 0.0, 0.2, 0.4, 0.6 and 0.8 V. The isographs are generated using the Inelastica code.\cite{inelastica} At zero bias voltage, it is obvious that the two spin-up HOMO and HOMO-1 are delocalized across the Fe(II)S complex as well as strongly coupling Fe(II)S to the two electrodes, whereas the spin-down frontier orbitals HOMO and LUMO are more localized to the left and right electrode, respectively, while being weakly localized at the opposite electrode. In addition, the separation between the spin-down frontier orbitals is much larger than that in the spin-up frontier orbitals (0.897 eV compared to 0.376 eV). As the voltage bias is increased, one can observe that the localization characteristics of the spin-down frontier orbitals are preserved until 0.6 V, where the LUMO becomes more localized at the left electrode. The MPSH eigenenergies of both the spin-down LUMO and HOMO are outside of the bias window, leading to weak conductance in the spin-down channel. On the other hand, in the spin-up channel, while the HOMO is always outside of the bias window for bias voltages $V_{\textrm{\tiny{\rm bias}}}=0.2$ V and 0.4 V, the LUMO is inside the bias window. It is also delocalized across the Fe(II)S complex, strongly coupled with both electrodes. Therefore, it is responsible for the large conductance in the spin-up channel. Interestingly, above the voltage bias $V_{\textrm{\tiny{\rm bias}}}=0.4$ V, the LUMO orbital ceases to contribute to transmission while the HOMO orbital dominates transmission. The NDR behavior is accompanied by the dominance of the LUMO in the spin-up channel for 0.2 and 0.4 V, and the dominance of the HOMO for 0.6 V and above. This shift in the conductive channel lead to the drop in current above 0.4 V, and can be understood by investigating the magnetic moment of the Fe atom as a function of bias voltage.

\begin{figure}
\includegraphics[width=90mm]{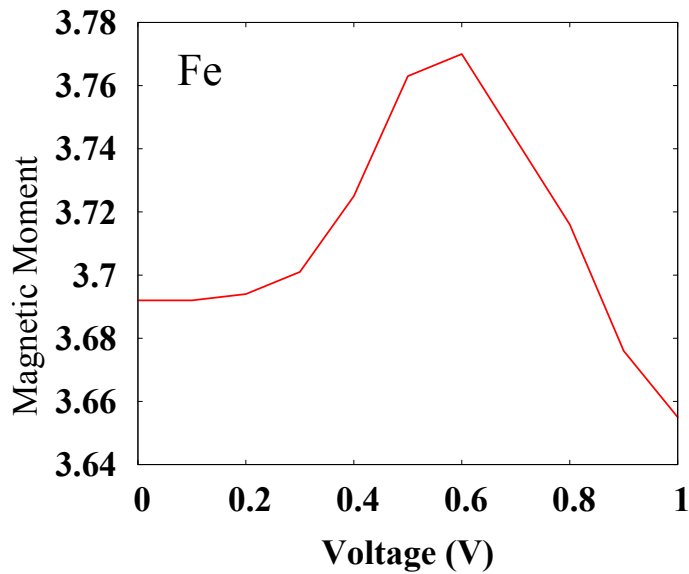}
  \caption{The Fe magnetic moment (in $\mu_B$) as a function of bias voltage $V_{\textrm{\tiny{\rm bias}}}$.}
  \label{fig:fig7}
\end{figure}

Figure~\ref{fig:fig7} displays the change in the magnetic moment of the Fe atom (in $\mu_B$) as a function of $V_{\textrm{\tiny{\rm bias}}}$. The graph shows a trend that somewhat resembles that of the IVC: a rise in magnetic moment up to 0.6 V, followed by a fall. Although the difference between the highest and lowest magnetic moments is quite small ($\sim 0.1 \mu_B$), such small change in moment can be viewed in light of the MPSH orbital analysis in Fig. \ref{fig:fig5}. When a potential difference is applied to the gold-Fe(II)S-gold system, the Fe atom gains a small amount of spin-up electron density. This additional spin-up spin density occupies the MPSH LUMO (recall that, as $V_{\textrm{\tiny{\rm bias}}}$ increases from 0.2 to 0.4 V, the LUMO is the conductive channel in the spin-up channel). Then, the magnetic moment of Fe starts to drop at 0.6 V (which corresponds to Fe electrons moving from the LUMO back to the HOMO), and the MPSH HOMO is now the conductive channel. Therefore, while the presence of electric potential slightly affects the magnetic moment of the Fe atom, such slight effect has significant consequences in spin-polarized transport.

In conclusion, the present investigation reports exceptionally high spin filtering efficiency and negative differential resistance of the nanodevice comprised of the high-spin state of a Fe(II)S molecule between two gold electrodes. The analysis of the transmission eigenchannels of the Au-Fe(II)S-Au system suggests that the rise in the spin-up conductivity up to a the bias voltage of 0.4 V is dominated by a conductive MPSH LUMO, and this is accompanied by a slight increase in the magnetic moment of the Fe atom. The subsequent drop in the spin-up conductivity is because the conductive channel is the MPSH HOMO, which has a low conductance contribution. This is accompanied by a drop in the magnetic moment of the Fe atom. The low-voltage negative differential resistance and the exceptionally high spin-filtering efficiency shows that the Fe(II)S molecular magnet can be efficiently implemented in spintronic applications.

This research was undertaken with the assistance of resources from the National Computational Infrastructure (NCI), which is supported by the Australian Government.


\end{document}